\documentclass[preprin,showpacs,preprintnumbers,eqsecnum,amsmath,amssymb,nofootinbib]{revtex4-1}
\usepackage{graphics,epsfig}
\usepackage{graphicx}
\usepackage{dcolumn}
\usepackage{bm}

\begin{document}
\baselineskip=0.8 cm
\title{ Charged Einstein-aether black holes and Smarr formula}

\author{Chikun Ding${}^{a, b}$ }
\email{Chikun_Ding@baylor.edu}
\author{ Anzhong Wang$^{b, c, d}$\footnote{The corresponding author}}
 \email{Anzhong_Wang@baylor.edu}
 \author{ Xinwen Wang$^{b, c}$}
 \email{Xinwen_Wang@baylor.edu}
  \affiliation{$^{a}$ Department of Information Science and Technology, Hunan University of Humanities, Science and Technology, Loudi, Hunan
417000, P. R. China\\
$^{b}$ GCAP-CASPER, Physics Department, Baylor University, Waco, Texas 76798-7316, USA\\
$^{c}$  Institute  for Advanced Physics $\&$ Mathematics, Zhejiang University of Technology, Hangzhou 310032,  China\\ 
$^{d}$ Departamento de F\'{\i}sica Te\'orica, Instituto de F\'{\i}sica, UERJ, 20550-900, Rio de Janeiro, Brazil}

\vspace*{0.2cm}
\begin{abstract}
\baselineskip=0.6 cm
\begin{center}
{\bf Abstract}
\end{center}

In the framework of the Einstein-Maxwell-aether theory, we present two new classes of exact charged
black hole solutions, which are asymptotically flat and possess the universal as well as Killing horizons.
We also construct the Smarr formulas, and calculate the temperatures of the horizons, using the Smarr
mass-area relation. We find that, in contrast to the neutral  case, such obtained temperature is not
proportional to its surface gravity at any of the two kinds of  the horizons. Einstein-Maxwell-aether black 
holes with the cosmological constant and their topological cousins  are also given.


\end{abstract}

 \maketitle

\vspace*{0.2cm}
\section{Introduction}

Lorentz invariance is one of the fundamental principles of Einstein's general relativity (GR) and modern physics. The success of GR to describe all observed gravitational phenomena, together with its intrinsic mathematical elegance is interpreted as a further proof of the  importance of the Lorentz invariance  \cite{blas}. However, Lorentz invariance may not be an exact symmetry at all energies  \cite{mattingly}. Any effective description must break down at a certain cutoff scale signaling the emergence of new physical degrees of freedom beyond that scale. For example, the hydrodynamics, Fermi's theory of beta decay \cite{bhattacharyya2} and quantization of GR \cite{shomer} at energies beyond the Planck energy. Astrophysical observations suggest that the high energy cosmic rays above the Greisen-Zatsepin-Kuzmin cutoff is a result of the Lorentz violation \cite{carroll}. Lorentz invariance also leads to divergences in quantum field theory which can be cured with a short distance of cutoff that breaks it \cite{jacobson}. Often, the late cosmic acceleration is also  interpreted as a demand for a modification of GR  at cosmological scales \cite{yagi,blas2}.

There are several gravitational theories that violate the Lorentz invariance  \cite{mattingly}, e.g., Ho\u{r}ava-Lifshitz theory \cite{Horava}, ghost condensations \cite{Shinji}, warped brane worlds,
Einstein-aether theory \cite{jacobson2},
etc. In Einstein-aether theory, the background tensor fields break the Lorentz symmetry, and were once thought must to be dynamical \cite{jacobson2},  but  more careful investigations  recently revealed  that it is not necessary  \cite{jacobson6}.
In this theory, the Lorentz symmetry is broken only down to a rotation subgroup by the existence of a preferred time direction at every point of spacetime, i.e., existing a preferred frame of reference established by aether vector $u^a$. This timelike unit vector field $u^a$ can be interpreted as a velocity four-vector of some medium substratum (aether, vacuum or dark fluid), bringing into consideration of non-uniformly moving continuous media and their interaction with other fields. Meanwhile, this theory can be
also considered as  a realization of dynamic self-interaction of complex systems moving with a spacetime dependant macroscopic velocity. As to an accelerated expansion of the universe, this dynamic self-interaction can produce the same cosmological effects as the dark energy \cite{balakin2}.

The introduction of the aether vector allows for some novel effects, e.g., matter fields can travel faster than the speed of light \cite{jacobson3}, new gravitational wave polarizations can spread at different speeds \cite{jacobson4}. It should be noted  that  the propagation faster than that of the light does not violate causality \cite{berglund}. In particular, gravitational theories with breaking Lorentz invariance still allow the existence of black holes \cite{blas2,barausse,lin,UHs}. However, instead of Killing horizons, now the boundaries of black holes are  hypersurfaces, termed
 universal horizons \cite{barausse,blas2},
which can trap excitations traveling at arbitrarily high velocities.
This universal horizon may radiate thermally at a fixed temperature and strengthen a possible thermodynamic interpretation though there is no universal light cone \cite{berglund2} (See also \cite{michel} for a different suggestion.).

It is naturally to extend the Einstein-aether theory to include other fields, i.e., electromagnetic one \cite{balakin}. As for cosmology, the interaction of electromagnetic waves with a non-uniformly moving aether can change the details of the standard history of the relic photons that could be tested using observational  data. As for black holes, the interaction of electromagnetic radiation with a deformed aether will induce new dynamo-optical effects that could be also tested. As for gravitational waves, the Einstein-Maxwell-aether theory is expected to predict new forms for gravitational wave propagations \cite{yagi,jacobson4}. Our goal here is to extend Einstein-aether theory to include a source-free Maxwell field. For more general formalism of the Einstein-Maxwell-aether theory,  see \cite{balakin}.

The rest of the paper is organized as follows. In Sec. II we provide the background for the Einstein-Maxwell-aether theory to be studied in this paper. In Sec. III we construct a Smarr formula for spherically symmetric solutions. In Sec. IV, we first construct  two new classes of exact charged solutions, and then use them as examples to study the Smarr formula. In Sec. V, we present our main conclusions.

Before proceeding further, we would like to note that the exact charged solutions presented in this paper can be considered as a generalization of the neutral ($Q = 0$) ones given  in \cite{berglund}. Therefore, in the following there may exist repeating of the materials presented there, in order  for the current paper to be as much independent as possible,  although  we shall try to limit this to its minimum. For more detail, we refer readers to \cite{berglund}.

\section{Einstein-Maxwell-aether theory}

The general action for the Einstein-aether theory can be constructed by assuming that: (1) it is general covariant; and (2)  it  is a functional of only the spacetime metric $g_{ab}$ and a unit timelike vector $u^a$, and involves  no more than two derivatives of them, so that the resulting field equations are second-order differential equations of  $g_{ab}$ and   $u^a$. Then,  the  Einstein-Maxwell-aether theory to be studied   in this paper is
 described by the  action,
\begin{eqnarray}
\mathcal{S}=
\int d^4x\sqrt{-g}\Big[\frac{1}{16\pi G_{\ae}}(\mathcal{R}+\mathcal{L}_{\ae})+\mathcal{L}_M\Big]\,. \label{action}
\end{eqnarray}
In terms of the tensor $Z^{ab}_{~~cd}$ defined as \cite{eling,garfinkle},
\begin{eqnarray}
Z^{ab}_{~~cd}=c_1g^{ab}g_{cd}+c_2\delta^a_{~c}\delta^b_{~d}
+c_3\delta^a_{~d}\delta^b_{~c}-c_4u^au^bg_{cd}\,,
\end{eqnarray}
 the aether Lagrangian $\mathcal{L}_{\ae}$ is given by
\begin{eqnarray}
-\mathcal{L}_{\ae}=Z^{ab}_{~~cd}(\nabla_au^c)(\nabla_bu^d)-\lambda(u^2+1),
\end{eqnarray}
where $c_i (i = 1, 2, 3, 4)$ are coupling constants of the theory.
The aether Lagrangian is therefore the sum of all possible terms for the aether field $u^a$ up to mass dimension two, and the constraint term $\lambda(u^2 + 1)$ with the Lagrange multiplier $\lambda$ implementing the normalization condition $u^2=-1$.
The source-free Maxwell Lagrangian $\mathcal{L}_M$ is given by
\begin{equation}
\mathcal{L}_M=-\frac{1}{16\pi G_{\ae}}\mathcal{F}_{ab}\mathcal{F}^{ab},
~\mathcal{F}_{ab}=\nabla_a\mathcal{A}_b-\nabla_b\mathcal{A}_a,
\end{equation}
where $\mathcal{A}_a$ is the electromagnetic potential four-vector.

There are a number of theoretical and observational bounds on the coupling constants $c_i$ \cite{jacobson2,yagi,jacobson5}. Here,  we impose the following
constraints\footnote{Note the slight difference between the constraints imposed here and the ones imposed in \cite{berglund}, as in this paper we also require that
vacuum Cerenkov radiation of gravitons is forbidden \cite{EMS}.},
\begin{equation}
\label{CDs}
0\leq c_{14}<2,\quad 2+c_{13}+3c_2>0,\quad 0\leq c_{13}<1,
\end{equation}
where $c_{14}\equiv c_1+c_4$, and so on. The constant $G_{\ae}$ is related to Newton's gravitational constant $G_{N}$ by $G_{\ae}=(1-c_{14}/2)G_N$, which can be
obtained by using the weak field/slow-motion limit of the Einstein-aether theory \cite{carroll,eling}.

The equations of motion, obtained by varying  the action (\ref{action}) with respect to $g_{ab}$, $u^a$, $\mathcal{A}^a$ and $\lambda$ are
\begin{eqnarray}\label{motion}
\mathcal{G}_{ab}=\mathcal{T}^{\ae}_{ab}+8\pi G_{\ae}\mathcal{T}^M_{ab},\quad {\AE}_a=0,\quad \nabla ^a\mathcal{F}_{ab}=0, \quad u^2=-1,
\end{eqnarray}
respectively, where the aether and Maxwell energy-momentum stress tensors $\mathcal{T}^{\ae}_{ab}$ and $\mathcal{T}^M_{ab}$ are given by
\begin{eqnarray}
\label{EMTs}
&&\mathcal{T}^{\ae}_{ab}=\lambda u_au_b+c_4a_aa_b-\frac{1}{2}g_{ab}Y^c_{~~d}\nabla_cu^d+\nabla_cX^c_{~~ab}
+c_1[(\nabla_au_c)(\nabla_bu^c)-(\nabla^cu_a)(\nabla_cu_b)],\nonumber \\
&&\mathcal{T}_{ab}^M
 =\frac{1}{4\pi G_{\ae}}\Big[-\frac{1}{4}g_{ab}\mathcal{F}_{mn}\mathcal{F}^{mn}
 +\mathcal{F}_{am}\mathcal{F}_{b}^{~m}\Big],
 \end{eqnarray}
with
\begin{eqnarray}
{\AE}_a=\nabla_bY^b_{~~a}+\lambda u_a+c_4(\nabla_au^b)a_b,\quad
Y^a_{~~b}=Z^{ac}_{~~~bd}\nabla_cu^d, \quad
 X^c_{~~ab}=Y^c_{~~(a}u_{b)}-u_{(a}Y^{~~c}_{b)}+u^cY_{(ab)} .
\end{eqnarray}
The acceleration vector $a^a$ appearing in the expression for the aether energy-momentum stress tensor is defined as the parallel transport of the aether field along itself, $a^a
\equiv \nabla_uu^a,$ where $\nabla_X\equiv X^b\nabla_b$.

Following \cite{berglund},  we first define a set of basis vectors at every point in the spacetime,
so that we can project out various components of the equations of motion.
Let us first  take the aether field $u^a$ to be the basis vector. Then, pick up two spacelike unit vectors, denoted, respectively, by $m^a$ and $n^a$, both of which are normalized to unity,  mutually orthogonal, and lie on the tangent plane of the two-spheres $\mathcal{B}$ that foliate the hypersurface $\Sigma_U$. Finally, let us pick up $s^a$, a spacelike unit vector that is orthogonal to $u^a$, $m^a$, $n^a$, and  points ``outwards'' along a $\Sigma_U$ hypersurface, so we have the four tetrad, $
e^a_{(b)} \equiv \left(u^a, s^a, m^a, n^a\right)$,  with
\begin{eqnarray}
 g^{ab} = \eta^{cd}e^a_{(c)} e^{b}_{(d)} = -u^au^b + s^as^b + \hat{g}^{ab}, \quad e_{(b)}\cdot e_{(c)} = \eta_{bc},
\end{eqnarray}
 where $\hat{g}^{ab} \equiv m^am^b + n^an^b$.
 By spherical symmetry, any physical vector $A^a$ has at most two non-vanishing components along, respectively,  $u^a$ and $s^a$, i.e., $A^a=A_1u^a+A_2s^a$.  In particular,
 the acceleration $a^a$  has only one component along $s^a$, namely,  $a^a=(a\cdot s)s^a$.
Similarly, any rank-two tensor $F_{ab}$ may have components along the directions of the bi-vectors $u_au_b,~u_{(a}s_{b)},~u_{[a}s_{b]},~s_as_b,~\hat{g}_{ab}$, where $\hat{g}_{ab}$ is the projection tensor onto the two-sphere $\mathcal{B}$, bounding a section of a $\Sigma_U$ hypersurface.
In the following, we study the expansion of the Maxwell field $\mathcal{F}^{ab}$,  Killing vector $\chi^a$,  surface gravity $\kappa$,  energy-momentum stress  tensors $\mathcal{T}_{ab}^{\ae}$ and $\mathcal{T}_{ab}^{M}$, and Ricci tensor $\mathcal{R}_{ab}$.
The given source-free Maxwell field $\mathcal{F}^{ab}$ can be formulated in
terms of four-vectors representing physical fields. They are
 the electric field $E^a$ and  magnetic excitation $B^a$ as,
 \begin{eqnarray}
E^a=\mathcal{F}^{ab}u_b, \quad B^a=\frac{e^{abmn}}{2\sqrt{-g}}\mathcal{F}_{mn}u_b,
\end{eqnarray}
where $e^{abmn}$ is the Levi-Civita tensor. From Eq.(\ref{motion}) it can be shown
$B^a=0$. Then, we find
\begin{eqnarray}
\mathcal{F}^{ab}=-E^au^b+E^bu^a.
\end{eqnarray}
On the other hand,  the electric field is spacelike, since
$E^au_a=0$. So, we have $E^a=(E\cdot s)s^a$. Thus,  $\mathcal{F}_{ab}=(E\cdot s)(-s_au_b+s_bu_a)$. After substituting it into (\ref{motion}), we can see $(E\cdot s)=Q/r^2$, where $Q$ is an integral constant, representing the total charge of the space-time. Therefore, we have
\begin{eqnarray}
\label{Maxwellb}
\mathcal{F}_{ab}=\frac{Q}{r^2}(u_as_b-u_bs_a).
\end{eqnarray}

The Einstein,  aether  and Maxwell equations of motion (\ref{motion}) can be decomposed  by using the tetrad defined above.
In particular,  the aether and electromagnetic energy-momentum stress tensors and the Ricci tensor can be cast, respectively, in the forms,
\begin{eqnarray}\label{projection}
&&\mathcal{T}^{\ae}_{ab}=\mathcal{T}^{\ae}_{uu}u_au_b-2\mathcal{T}^{\ae}_{us}u_{(a}s_{b)}
+\mathcal{T}^{\ae}_{ss}s_as_b+\frac{\hat{\mathcal{T}}_{\ae}}{2}\hat{g}_{ab},\nonumber\\
&&\mathcal{R}_{ab}=\mathcal{R}_{uu}u_au_b-2\mathcal{R}_{us}u_{(a}s_{b)}
+\mathcal{R}_{ss}s_as_b+\frac{\mathcal{\hat{R}}}{2}\hat{g}_{ab},\nonumber\\
&&\mathcal{T}^{M}_{ab}=\mathcal{T}^{M}_{uu}u_au_b-2\mathcal{T}^{M}_{us}u_{(a}s_{b)}
+\mathcal{T}^{M}_{ss}s_as_b+\frac{\hat{\mathcal{T}}_{M}}{2}\hat{g}_{ab}.
\end{eqnarray}
The coefficients of $\mathcal{T}^{\ae}_{ab}$ and $\mathcal{T}^{M}_{ab}$ in (\ref{projection}) can be computed from the general expression (\ref{EMTs}). The corresponding coefficients for $\mathcal{R}_{ab}$, on the other hand, are computed from the definition $[\nabla_a,~\nabla_b]X^c \equiv -\mathcal{R}^c_{~abd}X^d$ by choosing $X^a = u^a$ or $s^a$, and then contracting the resulting expressions again with $u^a$ and/or $s^a$ appropriately. The coefficients for the  three $(u, s)$ cross terms are
\begin{eqnarray}\label{usT}
\mathcal{T}^{\ae}_{us}=c_{14}\left[\hat{K}(a\cdot s)+\nabla_u(a\cdot s)\right],\quad
\mathcal{T}^{M}_{us}=0,\quad \mathcal{R}_{us}=(K_0-\hat{K}/2)\hat{k}-\nabla_s\hat{K},
\end{eqnarray}
where
\begin{eqnarray}\label{usTa}
\nabla_{[a}s_{b]} \equiv - K_0 u_{[a}s_{b]},\;\;\;
\hat{k} \equiv \frac{1}{2} g^{ab} {\cal{L}}_s\hat g_{ab},\;\;\;
\hat{K} \equiv \frac{1}{2} g^{ab} {\cal{L}}_u\hat g_{ab},
\end{eqnarray}
with $K \; (\equiv K_0+\hat{K}) $ being  the trace of the extrinsic curvature of the  hypersurface $\Sigma_U$.
The aether equation $s\cdot{\AE}=0$ and   the $us$-component  $\mathcal{R}_{us}=\mathcal{T}^{\ae}_{us} + 8\pi G_{\ae} \mathcal{T}^{M}_{us}$ yield
\begin{eqnarray}\label{us}
 && c_{123}\nabla_s K_0-(1-c_{13})(K_0-\hat{K}/2)\hat{k}+(1+c_2)\nabla_s\hat{K}=0,\\
 \label{usB}
&&  c_{123}\nabla_s K-(1-c_{13})\mathcal{T}^{\ae}_{us} = 0.
\end{eqnarray}
The  $uu$- and $ss$-components of the gravitational field equations  give
\begin{eqnarray}\label{uuss}
\left(1-\frac{c_{14}}{2}\right)\left[(\hat{k}+\nabla_s)(a\cdot s)+a^2\right]-(1-c_{13})\left(K_0^2+\frac{\hat{K}^2}{2}\right)\nonumber\\
-\left(1+c_2+\frac{c_{123}}{2}\right)\nabla_uK
-\frac{c_{123}}{2}K^2-\frac{Q^2}{r^4}=0,\\
\label{uussB}
\frac{c_{123}}{2}(K+\nabla_u)(\hat{K}-K_0)+(1+c_2)KK_0-\left(1+\frac{c_{14}}{2}\right)
\left[\nabla_s(a\cdot s)+a^2\right]\nonumber\\
+c_{14}a^2-\left[\nabla_s+\frac{\hat{k}}{2}+\frac{c_{14}}{2}(a\cdot s)\right]\hat{k}+\frac{Q^2}{r^4}=0.
\end{eqnarray}
In the next sections, we will use these equations to obtain new  black holes solutions.

\section{Smarr formula}

The studies of black holes have been one of the main objects both theoretically and observationally over the last half of
century \cite{BHsOb,BHsTheor}, and so far there are many solid  observational evidences for their existence in our universe.
Theoretically,  such investigations   have been playing a fundamental   role in the understanding of the nature of  gravity in general, and quantum gravity in particular.
They  started with  the discovery of the laws of black hole mechanics \cite{BCH} and Hawking radiation \cite{HawkingR}, and   led to  the  profound
 recognition of the thermodynamic interpretation of the four laws \cite{Bekenstein73} and  the reconstruction of general relativity (GR) as the thermodynamic
 limit of a more fundamental theory of gravity \cite{Jacobson95}.  More recently, they are essential in understanding  the AdS/CFT correspondence
  \cite{tHS,MGKPW} and firewalls \cite{AMPS}.

To derive the Smarr formula, we first introduce the ADM mass, which is identical to the Komar mass defined in stationary spacetimes with the time translation Killing vector
$\chi^a$ \cite{berglund},
\begin{eqnarray}\label{KM}
M_{ADM} =-\frac{1}{4\pi G_{\ae}}\int_{{\cal{B}}_{\infty}}{\nabla^a\chi^b d\Sigma_{ab}},
\end{eqnarray}
where $d\Sigma_{ab} \equiv  - u_{[a}s_{b]}dA$, with $dA \; (\equiv r^2\sin\theta d\theta d\phi)$ being the differential area  element  on the two-sphere ${\cal{B}}$,
and ${\cal{B}}_{\infty}$ is the sphere at infinity.
The derivative of the Killing vector $\chi^a=-(u\cdot\chi)u^a+(s\cdot\chi)s^a$ is given by
\begin{eqnarray}\label{kappa}
\nabla^a\chi^b=-2\kappa u^{[a}s^{b]},\quad
\end{eqnarray}
where $\kappa$ denotes  the surface gravity usually defined in GR, and is given by
\begin{eqnarray}\label{kappa1}
\kappa =\sqrt{-\frac{1}{2}(\nabla_a\chi_b)(\nabla^a\chi^b)}=-(a\cdot s)(u\cdot\chi)+K_0(s\cdot\chi).
\end{eqnarray}
At the infinity, we have $(u\cdot\chi) = -1$ and $(s\cdot\chi) = 0$. Then, Eq.(\ref{KM}) yields,
\begin{eqnarray}\label{KMb}
M_{ADM} =\lim_{r\rightarrow \infty}\left(\frac{r^2(a\cdot s)}{G_{\ae}}\right).
\end{eqnarray}

In the studies of black hole physics, the physics of horizons has provided useful information.
In particular, at the Killing horizon  the first law \cite{smarr} and Smarr formula \cite{manko} for the Reissner-Nordstrom black hole take the forms,
\begin{eqnarray}\label{madm}
\delta M_{ADM}=\frac{\kappa_{KH}\delta A_{KH}}{8\pi G_N}+\frac{V_{KH}\delta Q}{ G_N},\quad
M_{ADM}=\frac{\kappa_{KH} A_{KH}}{4\pi G_N}+\frac{V_{KH} Q}{G_N},
\end{eqnarray}
where $M_{ADM}$ is the ADM mass of the spacetime and, $\kappa_{KH} [\equiv \kappa(r_{KH})]$, $A_{KH}$ and $V_{KH}$ are the surface gravity, cross-sectional area and electromagnetic potential evaluated on the Killing horizon, respectively. Identifying $T_{KH}=\kappa_{KH}/2\pi$ as the temperature of the horizon and, the entropy $S=A_{KH}/4G_N$, one can obtain the analogy with the first law of thermodynamics, $\delta E = T \delta S+V\delta Q$ and $E=2TS+VQ$.

Any causal boundary in a gravitational theory should have an entropy associated with it. Therefore,  in the Einstein-aether theory,  the universal horizons are expected to have also their entropy and the first law of black hole mechanics, though whether this entropy is to be proportional to its area or not is still an open issue. Meanwhile, these black holes still have Killing horizons. Then, a question is:   How is their thermodynamics?

To obtain some hints, in this section we shall present the Smarr formulas of the universal and Killing horizons for  general static and spherically symmetric Einstein-Maxwell-aether black holes. Let us first consider the geometric identity \cite{BCH},
\begin{eqnarray}
\label{ID}
\mathcal{R}_{ab}\chi^b=\nabla^b(\nabla_a\chi_b).
\end{eqnarray}
From the Einstein field equations (\ref{motion}), we find that $\mathcal{R}_{ac} = \mathcal{T}^{\ae}_{ab} - g_{ab}\mathcal{T}^{\ae}/2 + 8\pi G_{\ae} \mathcal{T}^M_{ab}$, where
\cite{bhattacharyya2}
\begin{eqnarray}
\label{IDa}
&& 8\pi G_{\ae} \mathcal{T}^M_{ab}\chi^b = - \frac{Q^2}{r^4}\chi_a,\nonumber\\
&&   \left(\mathcal{T}^{\ae}_{ab} - \frac{1}{2} g_{ab}\mathcal{T}^{\ae}\right)\chi^b = \nabla^b\Big\{\big[c_{14}\left(a\cdot s\right) \left(u\cdot \chi\right)
+ \left(c_{123} K  - 2c_{13} K_0\right) \left(a\cdot s\right)\big] u_{[a}s_{b]}\Big\}.
\end{eqnarray}
Setting
\begin{eqnarray}
\label{IDb}
- \frac{Q^2}{r^4}\chi_a = 2 \nabla^b\left(F^Q(r) u_{[a}s_{b]}\right),
\end{eqnarray}
we find that   Eq.(\ref{ID}) can be cast in the form,
\begin{eqnarray}
\label{SmarrFa}
\nabla_bF^{ab}=0,\;\;\; F^{ab} \equiv 2F(r)u^{[a}s^{b]}, 
\end{eqnarray}
where
\begin{eqnarray}
\label{SmarrFb}
&& F(r) =  F^Q(r) + q(r),\nonumber\\
&& q(r)  \equiv    -\left(1-\frac{c_{14}}{2}\right)(a\cdot s)(u\cdot \chi)+\left[(1-c_{13})K_0 +\frac{c_{123}}{2}K\right](s\cdot \chi).
\end{eqnarray}
On the other hand, comparing Eq.(\ref{SmarrFa}) with the soucre-free Maxwell equations  (\ref{motion}), we find that its solution must also take the form (\ref{Maxwellb}), that is,
$F(r) = F_0/r^2$. To determine the integration constant $F_0$, we note that, for asymptotically flat space-time, we have \cite{eling,berglund},
\begin{eqnarray}
\label{ASYMP}
 u\cdot\chi\simeq -1,\;\;\;\; s\cdot\chi\simeq 0,\;\;\;\;
 a\cdot s = \frac{r_0}{2r^2} + {\cal{O}}(r^{-3}),
 \end{eqnarray}
 as $r\rightarrow\infty$.   Then, from Eq.(\ref{SmarrFb}) we find that $F_0 = {r_0}(1-{c_{14}}/{2})/{2}$. Thus, we have
\begin{eqnarray}
F(r) =\left(1-\frac{c_{14}}{2}\right)\frac{r_0}{2r^2}.
\end{eqnarray}

Inserting Eq.(\ref{ASYMP}) into Eq.(\ref{KMb}), on the other hand,  we find that  the ADM mass  is given by $M_{ADM}=r_0/2G_{\ae}$.
Therefore,  the total mass $M$ of the spacetime is
\begin{eqnarray}\label{tmass}
M \equiv M_{ADM}+M_{\ae}  =\left(1-\frac{c_{14}}{2}\right)\frac{r_0}{2G_{\ae}} 
=\frac{1}{4\pi G_{\ae}}\int_{\mathcal{B}_\infty}FdA,
\end{eqnarray}
where $M_{\ae}=-c_{14}M_{ADM}/2$ is the aether mass or aether contribution to the renormalization of $M_{ADM}$ \cite{eling}.
On the other hand, using Gauss' law,  from Eq.(\ref{SmarrFa}) we find that
\begin{eqnarray}\label{tmassB}
   0 = \int_{\Sigma}{\left(\nabla_bF^{ab}\right)   d\Sigma_a} =   \int_{{\cal{B}}_{\infty}}{F^{ab}   d\Sigma_{ab}} - \int_{{\cal{B}}_{H}}{F^{ab}   d\Sigma_{ab}}
   =   \int_{{\cal{B}}_{\infty}}{FdA} - \int_{{\cal{B}}_{H}}{F dA}.
\end{eqnarray}
Here $d\Sigma_a$ is the surface element of a spacelike  hypersurface $\Sigma$. The boundary $\partial\Sigma$ of $\Sigma$
consists of the boundary  at spatial infinity ${\cal{B}}_{\infty}$,  and the horizon ${\cal{B}}_{H}$, either the Killing or the universal.
Note that Eq.(\ref{tmassB}) is nothing but the conservation law of the flux of $F^{ab}$. Substituting the above expression into Eq.(\ref{tmass})  and taking
Eq.(\ref{SmarrFb})  into account, we find the following Smarr formula,
\begin{eqnarray}
MG_{\ae}=\frac{q_{UH}A_{UH}}{4\pi}+V_{UH}Q,\quad MG_{\ae}=\frac{q_{KH}A_{KH}}{4\pi}+V_{KH}Q,
\end{eqnarray}
where $A_{UH}$ and $A_{KH}$ are, respectively,  the area of the universal and Killing horizons, and $M$ is the total mass of an asymptotically flat solution defined in the asymptotic aether rest frame.
The potential $V_H$ is defined as $V_H\equiv r^2_HF^Q(r_H)/Q$.
Hence, the first law for the aether black hole may be obtained via a variation of the Smarr relation.
In the next section we consider it for  two new classes of exact charged aether black hole solutions.

For  the surface gravity at the universal horizon, when one  considers the peeling behavior of particles moving at any speed, i.e., capturing the role of the aether in the propagation of the physical rays, one finds that the  surface gravity at the universal horizon  is \cite{cropp,jacobson5,lin}
\begin{eqnarray}\label{kappa}
\kappa_{UH}\equiv \frac{1}{2}\nabla_u(u\cdot\chi)
= \left.\frac{1}{2}\left(a\cdot s\right)\left(s\cdot \chi\right) \right|_{r= r_{UH}},
\end{eqnarray}
where in the last step we used the fact that $\chi_a$ is a Killing vector,  $\nabla_{(a}\chi_{b)} =0$.  It must be noted that this is different from the surface gravity defined in GR  by Eq.(\ref{kappa1}). In particular, at the universal horizon we have $u\cdot \chi = 0$, and Eq.(\ref{kappa1}) yields,
\begin{eqnarray}\label{kappa2}
\kappa\left(r_{UH}\right) = \left. K_0(s\cdot\chi)\right|_{r= r_{UH}}.
\end{eqnarray}

\section{Exact Solutions of charged aether black holes}

To construct exact solutions of charged aether black holes, let us first choose the Eddington-Finklestein coordinate system, in which
the    metric takes the form
\begin{eqnarray}\label{metric}
ds^2=-e(r)dv^2+2f(r)dvdr+r^2d\Omega^2_2,
\end{eqnarray}
and the corresponding timelike Killing  and aether vectors are
\begin{equation}
  \chi^a=(1,0,0,0),\quad u^a=\big(\alpha,~\beta,~0,~0,\big),\quad u_adx^a=\big(-e\alpha+f\beta,~f\alpha,~0,~0,\big)\left(
\begin{array}{c}
 dv\\
 dr\\ d\theta\\ d\phi
\end{array}
\right),
\end{equation}
where $\alpha(r)$ and $\beta(r)$ are  functions of $r$ only. Then,  the metric can be written as  $g_{ab}=-u_au_b+s_as_b+\hat{g}_{ab}$, where we have the
 constraints $u^2=-1, ~s^2=1,~ u\cdot s=0$.
The boundary conditions on the metric coefficients are such that the solution is asymptotically flat, while those for the aether components are such that
\begin{eqnarray}\label{alpha}\lim_{r \rightarrow \infty} u^a = \{1, 0, 0, 0\}~.\end{eqnarray}

Some quantities that explicitly appear in Eqs.(\ref{us})-(\ref{uussB}) are  \cite{bhattacharyya}
\begin{eqnarray}
\label{AS}
 (a\cdot s)=-\frac{(u\cdot \chi)'}{f},\quad K_0=-\frac{(s\cdot \chi)'}{f},\quad\hat{K}=-\frac{2(s\cdot \chi)}{rf},\quad\hat{k}=-\frac{2(u\cdot \chi)}{rf},
\end{eqnarray}
where a prime $(')$ denotes a derivative with respect to $r$. And $\alpha(r),~\beta(r)$ and $ e(r)$ are
\begin{eqnarray}\label{abe}
 \alpha(r)=\frac{1}{(s\cdot\chi)-(u\cdot\chi)},\quad \beta(r)=-\frac{(s\cdot\chi)}{f}, \quad e(r)=(u\cdot\chi)^2-(s\cdot\chi)^2.
\end{eqnarray}
Then, from Eqs.(\ref{kappa}) and (\ref{kappa2}) we obtain
\begin{eqnarray}\label{abeA}
 \kappa_{UH} =-\left. \frac{1}{2f}\left(u\cdot \chi\right)' (s\cdot\chi)\right|_{UH},\quad
  \kappa(r_{UH}) =-\left. \frac{1}{f}\left(s\cdot \chi\right)' (s\cdot\chi)\right|_{UH}.
\end{eqnarray}
Clearly, in general $ \kappa_{UH} \not= \kappa(r_{UH})$.

From the above expressions one can see that all quantities can be calculated  from  $(u\cdot\chi)$ and $(s\cdot\chi)$ under the condition $f(r)=1$.
A straightforward  calculation of Eq.(\ref{usT}) yields
 \begin{eqnarray}
 \mathcal{R}_{us}=\frac{2(s\cdot\chi)(u\cdot\chi)f'(r)}{rf^3(r)}.
\end{eqnarray}
In the static spherical symmetric and asymptotically flat spacetime, if we assume that $f=1$ holds in the whole space-time,  we find
\begin{eqnarray}\label{abeA}
\mathcal{R}_{us}=\mathcal{T}^{\ae}_{us}=0,\; (f = 1).
\end{eqnarray}
From Eq.(\ref{IDb}), we also find that
\begin{eqnarray}\label{abeD}
F^Q(r)=- \frac{Q^2}{r^2}\int^{r}{\frac{f(r')}{{r'}^2}dr'} = \frac{Q^2}{r^3},\; (f =1). 
\end{eqnarray}

In the following,  we shall use the above expressions  first  to obtain two classes of exact solutions for the cases $c_{14}=0,~c_{123}\neq0$ and $c_{123}=0,~c_{14}\neq0$,
all with $f = 1$. Then, we shall study their main properties by using the Smarr formulas given above.

\subsection{Exact solutions for $c_{14}=0$}

When the coupling constant $c_{14}$ is set to zero and $c_{123}\neq 0$, from Eqs.(\ref{usB}) and (\ref{abeA}) one can see the quantity $\nabla_sK$ has to be vanished, i.e., $\nabla_sK=0$. So, the trace of the extrinsic curvature $K$ of the $\Sigma_U$ hypersurface is  constant. In the infinity, this constant will vanish asymptotically due to the asymptotical flat conditions. Therefore, it must vanish  everywhere.  Substituting $K=0$ into  Eqs.(\ref{uuss}) and (\ref{uussB}), we obtain
\begin{eqnarray}\label{aether}
 &&(s\cdot\chi)=\frac{r^2_{\ae}}{r^2},\nonumber\\
 &&(u\cdot\chi)=-\sqrt{1-\frac{r_0}{r}+\frac{Q^2}{r^2}+\frac{(1-c_{13})
 r^4_{\ae}}{r^4}},
\end{eqnarray}
where $r_{\ae}$ is another integral constant.
Then, using the formula (\ref{abe}), we find
\begin{eqnarray}\label{rn1}
 e(r)=1-\frac{r_0}{r}+\frac{Q^2}{r^2}-\frac{c_{13}r^4_{\ae}}{r^4},\quad f(r)=1,\end{eqnarray}
which reduces to those given  in Ref.  \cite{berglund} when $Q=0$.

\begin{figure}[ht]
\centering
\includegraphics[width=7.cm]{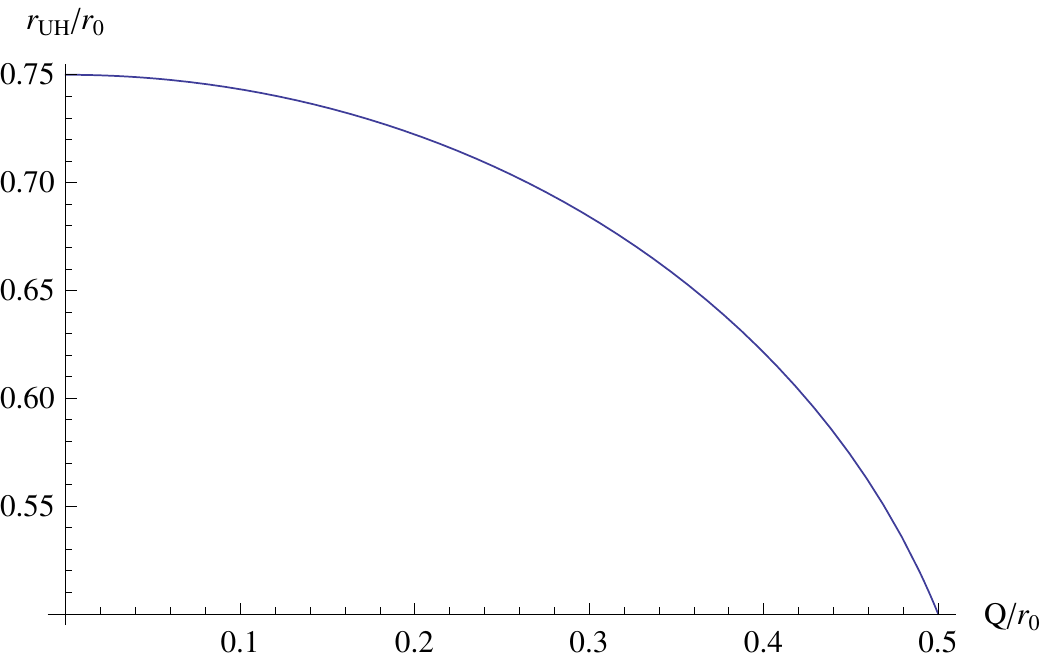}\;\;\;\;\;\includegraphics
[width=7.5cm]{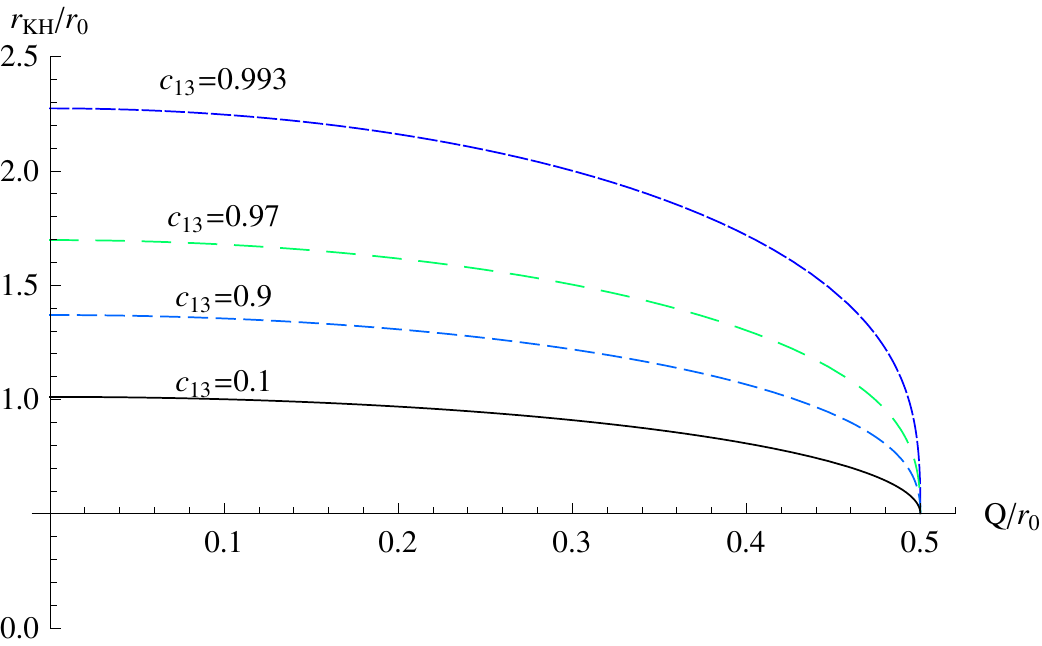}\;\;\;\;\;
\caption{The universal  and Killing horizons of the charged aether black hole with different $c_{13}$ in the case of $c_{14}=0,~c_{123}\neq0$. The presence of the charge $Q$ makes both  horizons smaller. The universal horizon does not depend on $c_{13}$, while the Killing horizon becomes bigger with the increasing of $c_{13}$. When $c_{13}=0$, the Killing horizon reduces to that of the Reissner-Nordstrom black hole.}
 \label{fig1}
  \end{figure}

The location of the universal horizon $r_{UH}$ is the largest root of equation $u\cdot\chi=0$. Meanwhile, $u\cdot\chi$ is a physical component of the aether, and should be regular and real everywhere. However,  from Eq.(\ref{aether}) one can see that in the region $r_{-}<r<r_{UH}$, this term becomes purely imaginary,
where $r_{-}$ is another root of $u\cdot\chi=0$, unless the two real roots coincide. Then, $r_{\ae}$ becomes a function of $r_0$. That is, the global existence of the
aether reduces the number of three independent constants ($r_0, r_{\ae}, Q$) to two,  ($r_0, Q$), the same as in GR.  Thus,  from  $(u\cdot\chi)^2 = 0$ and $d(u\cdot\chi)^2/dr = 0$  \cite{lin}, we find
\begin{eqnarray}\label{ruh1}
 r_{UH}=\frac{r_0}{2}\left(\frac{3}{4}+\sqrt{\frac{9}{16}-2\frac{Q^2}{r_0^2}}\right),\quad
 r^4_{\ae}=\frac{1}{1-c_{13}}\left(r_{UH}^4-\frac{1}{2}r_0r_{UH}^3\right),
\end{eqnarray}
which is showed in Figure \ref{fig1}.
One can see that the charge $Q$ is subjected to the condition
$ Q\leq3r_0/4\sqrt{2}$, in order to have $r_{UH}$ real.
When $Q=r_0/2$, we find $r_{\ae}=0$ and $r_{UH}=r_{KH}=r_0/2$. When $Q>r_0/2$, we have $r_{\ae}^4<0$. Thus, in order to have the aether be regular everywhere, the charge should be,
\begin{eqnarray}
 Q\leq\frac{r_0}{2},
\end{eqnarray}
which is the same as  that given in the Reissner-Nordstrom black hole.

\begin{figure}[ht]
\centering
\includegraphics[width=7.cm]{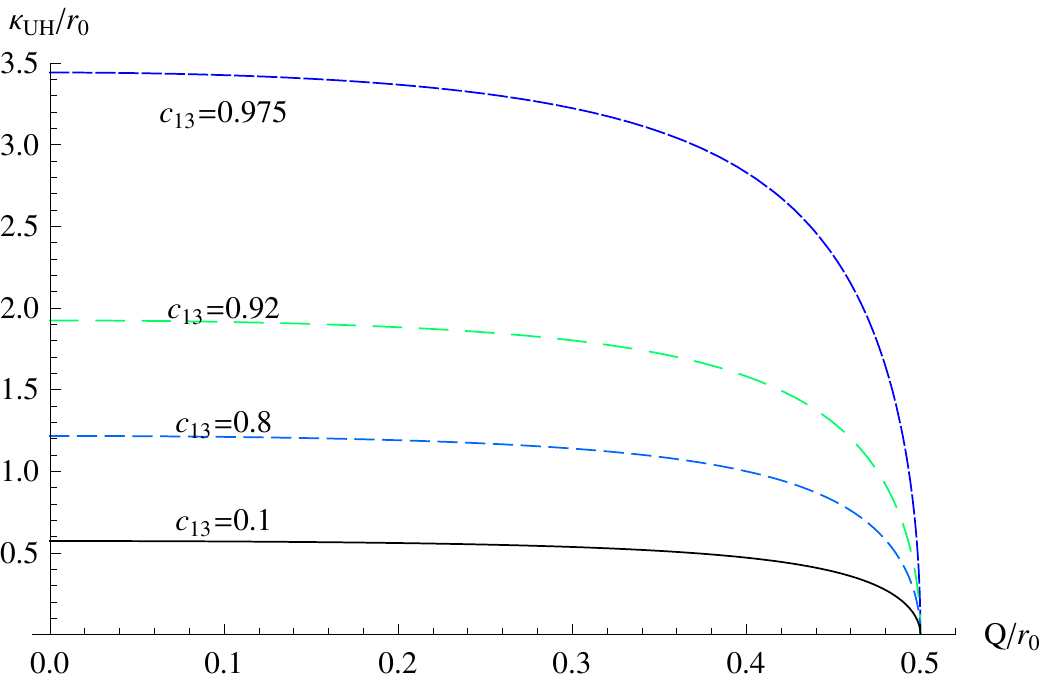}\;\;\;\;\;\includegraphics
[width=7.5cm]{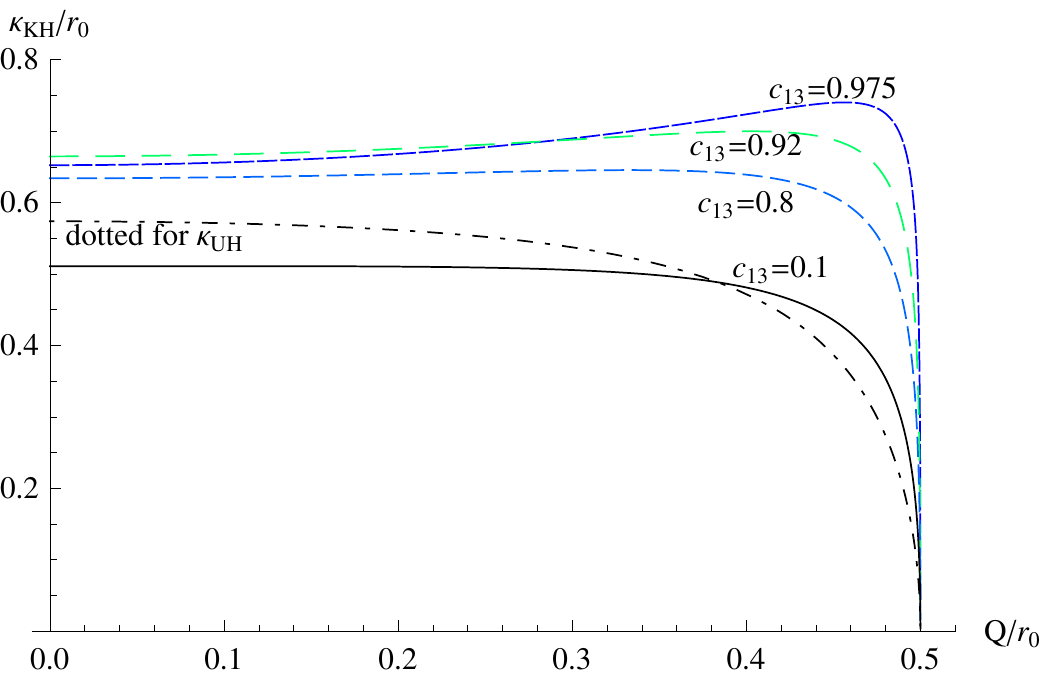}\;\;\;\;\;
\caption{The surface gravity at the universal  and Killing horizons of  the charged aether black hole in the case  $c_{14}=0,~c_{123}\neq0$. To compare with the results
given in Ref. \cite{lin}, we shift the $\kappa_{UH}$ line with $c_{13}=0.1$ from the left-hand side as a dot-dashed line to the right. One can see that when $c_{13}$ is small,  $\kappa_{UH}$ is larger  than $\kappa_{KH}$ in the low charge region, while lower in the large charge region, similar to that given  in Ref.  \cite{lin} for the
 Khronometric theory. }
 \label{fig2}
  \end{figure}

Now let us derive the Smarr relation.  Using Eq.(\ref{kappa}), the surface gravity at the universal horizon can be computed and is given by
\begin{eqnarray}\label{kappauh1}
 \kappa_{UH}=\frac{1}{2}\nabla_u(u\cdot\chi)=\frac{1}{2r_{UH}}
 \sqrt{\frac{2}{3(1-c_{13})}\left(1-\frac{Q^2}{2r^2_{UH}}\right)
 \left(1-\frac{Q^2}{r^2_{UH}}\right)},
\end{eqnarray}
which is showed in Figure \ref{fig2}. When $Q=0$, we find that $r_{UH}=3r_0/4$ and $\kappa_{UH}=\frac{2}{3r_0}\sqrt{\frac{2}{3(1-c_{13})}}$, which is the same as those given in  \cite{berglund2,cropp}.
The Smarr formula at the universal horizon is
\begin{eqnarray}\label{quh1}
 MG_{\ae}=\frac{q_{UH}A_{UH}}{4\pi}+V_{UH}Q,\quad q_{UH}=\frac{2}{3}\Big(\frac{1}
 {r_{UH}}-\frac{Q^2}{r_{UH}^3}\Big), \quad V_{UH} = \frac{Q}{r_{UH}},
\end{eqnarray}
  which don't depend explicitly on the coupling constants $c_i$'s,  because now we have $c_{14}=0$ and $M_{\ae}=0$.
It is easy to see that $q_{UH}$ isn't proportional to $\kappa_{UH}$ given by Eq.(\ref{kappauh1}).
On the other hand, at the universal horizon we find
\begin{eqnarray}\label{delta1}
 G_{\ae} \delta M=\frac{1}{8\pi}\left(\frac{2}{3r_{UH}}
  -\frac{Q^2}{3r_{UH}^3}\right)
  \delta A_{UH}+\frac{2}{3}V_{UH}\delta Q.
\end{eqnarray}
Why is there the factor $2/3$ in the front of $V_{UH}$? For a better understanding, let us use the method proposed in  \cite{smarr}, i.e., using $M$'s expression from $A=4\pi r_{UH}^2$,
\begin{eqnarray}
\label{delta1a}
  G_{\ae} M=\frac{1}{3}\sqrt{\frac{A}{\pi}+4Q^2+\frac{4\pi}{A}Q^4},
\end{eqnarray}
and writing the variation of $M$ as $G_{\ae} \delta M=T\delta A+V\delta Q$, we  obtain,
\begin{eqnarray}
\label{delta2}
  T\equiv \frac{\partial \left(G_{\ae} M\right)}{\partial A}=\frac{1}{8\pi}\left(\frac{2}{3r_{UH}}
  -\frac{Q^2}{3r_{UH}^3}\right),\quad V\equiv\frac{\partial \left(G_{\ae} M\right)}{\partial Q}=\frac{2Q}{3r_{UH}}=\frac{2}{3}V_{UH},
\end{eqnarray}
which are the same as those given in Eq.(\ref{delta1}). However, such defined temperature  $T$  is also not proportional to $\kappa_{UH}$ given by  Eq.(\ref{kappauh1}).

On the other hand, the location of the Killing horizon is the largest root of $e(r)=0$. Using Eq.(\ref{rn1}), we find
\begin{eqnarray}
 &&r_{KH}=\frac{r_0}{2}\left(\frac{1}{2}+L
 +\sqrt{N-P+\frac{1-4Q^2/r_0^2}{4L}}\right),\quad L=\sqrt{\frac{N}{2}+P},\quad
 N=\frac{1}{2}-\frac{4Q^2}{3r_0^2},\nonumber\\
 && P=\frac{2^{1/3}(12I+Q^4/r_0^4)}{3H}+\frac{H}{3\cdot2^{1/3}},\quad
 I=-\frac{c_{13}}{1-c_{13}}\left(\frac{r_{UH}^4}{r_0^4}
 -\frac{r_{UH}^3}{2r_0^3}\right),\nonumber\\
 &&
 H=\left(J+\sqrt{-4(12I+Q^4/r_0^4)^3+J^2}\right)^{1/3},\quad J=27I-72IQ^2/r_0^2
 +2Q^6/r_0^6,
\end{eqnarray}
which is showed in Figure \ref{fig1}. When $c_{13}=0$, we find that $r_{KH}=r_+$, that is, it coincides with the Reissner-Nordstrom black hole Killing horizon (here after we denote $r_{\pm} =(1\pm \sqrt{1-4Q^2/r_0^2})r_0/2.$).
Then,  the Smarr formula and surface gravity (using Eq.(\ref{kappa1})) at the Killing horizon are
\begin{eqnarray}\label{qkh1kappakh1}
 MG_{\ae}=\frac{q_{KH}A_{KH}}{4\pi }+V_{KH}Q,\quad q_{KH}=\left(\frac{r_0}{2r^2_{KH}}-\frac{Q^2}{r^3_{KH}}\right),
 \quad \kappa_{KH}=\frac{2}{r_{KH}}-\frac{3r_0}{2r_{KH}^2}+\frac{Q^2}{r_{KH}^3},
\end{eqnarray}
 which is showed in Figure \ref{fig2}.
 {Note again that  the $q_{KH}$ is still not proportional to the $\kappa_{KH}$.} When $c_{13}=0$, both $q_{KH}$ and
 $\kappa_{KH}$ reduce to those given in the Reissner-Nordstrom black hole,
  $q_{KH}=\kappa_{KH}=(r_+-r_-)/2r_+^2$. The first law at the Killing
horizon cannot be obtained via the variation method, although it may
be obtained via Smarr's method \cite{smarr}.
However, due to its complexity, we shall not consider this possibility, as even we do it, we do not expect to get much from such complicated expressions.

%

Note that,  when  $c_{13}\ll1$, from Eq.(\ref{rn1}) we find that  the solution reduces to  the usual Reissner-Nordstrom black hole with a universal horizon given by  (\ref{ruh1}) that is always inside its Killing horizon $r_{EH}=r_+$, which is the same as that derived in
the Khronometric theory  \cite{lin}.

Finally, let us turn to Figure \ref{fig2}, from which we can see that
the presence of the charge $Q$ always makes the { surface gravity}
$\kappa_{UH} $ lower, while the presence of the constant $c_{13}$ always makes it bigger,
after  the constraints (\ref{CDs}) are taken into account. For the
$\kappa_{KH}$, the situation becomes more complicated. In particular, when
 both   $c_{13}$ and $Q$ are small, the effects of them is similar to that
 presented in  $\kappa_{UH}$
  as  shown in the figure. But for  large $c_{13}$,
  e.g. $c_{13}=0.92,~0.975$, the presence of the charge increases the
   temperature at the beginning and then decreases it when the charge becomes
    very large.

\subsection{Exact solutions for $c_{123}=0$}

In this case, from Eq.(\ref{tmass}) we find that  the total mass is
\begin{eqnarray}
 MG_{\ae}=\left(1-\frac{c_{14}}{2}\right)\frac{r_0}{2}.
\end{eqnarray}
There exists a range of the coupling constants  that passes all the current observational tests in the one-parameter family of the Einstein-aether theories \cite{foster}.
Setting  $c_{123} = 0$,
 from  Eqs.(\ref{uuss}),  (\ref{uussB}) and (\ref{AS}) we obtain
\begin{eqnarray}
 &&(u\cdot\chi)=-1+\frac{r_0}{2r},\nonumber\\
 &&(s\cdot\chi)=\frac{r_0+2r_u}{2r},\quad r_u=\frac{r_0}{2}\left(\sqrt{\frac{2-c_{14}}{2(1-c_{13})}-\frac{4Q^2}
 {(1-c_{13})r_0^2}}-1\right).
\end{eqnarray}
Then, we find that
\begin{eqnarray}\label{rn2}
 e(r)=1-\frac{r_0}{r}-\frac{r_u(r_0+r_u)}{r^2},\;\;\; f(r) = 1,
\end{eqnarray}
which again reduces to  that given in Ref.  \cite{berglund} when $Q=0$.
From the above expressions, we find
\begin{eqnarray}
 \alpha(r)=\frac{1}{(s\cdot\chi)-(u\cdot\chi)}=\frac{1}{1+\frac{r_u}{r}}.
\end{eqnarray}
Since  it is one of the component of $u^a$, it should be regular everywhere (possibly except at the singular point $r = 0$), we must have
\begin{eqnarray}\label{alruh}
r_u\geq0\Rightarrow Q\leq \sqrt{\frac{2c_{13}-c_{14}}{2}}\; \frac{r_0}{2}
=\sqrt{1-\frac{c_{14}}{2}-(1-c_{13})}\;\;
\frac{r_0}{2},\quad c_{13}\geq \frac{c_{14}}{2}.
\end{eqnarray}

\begin{figure}[ht]
\centering
\includegraphics[width=7.cm]{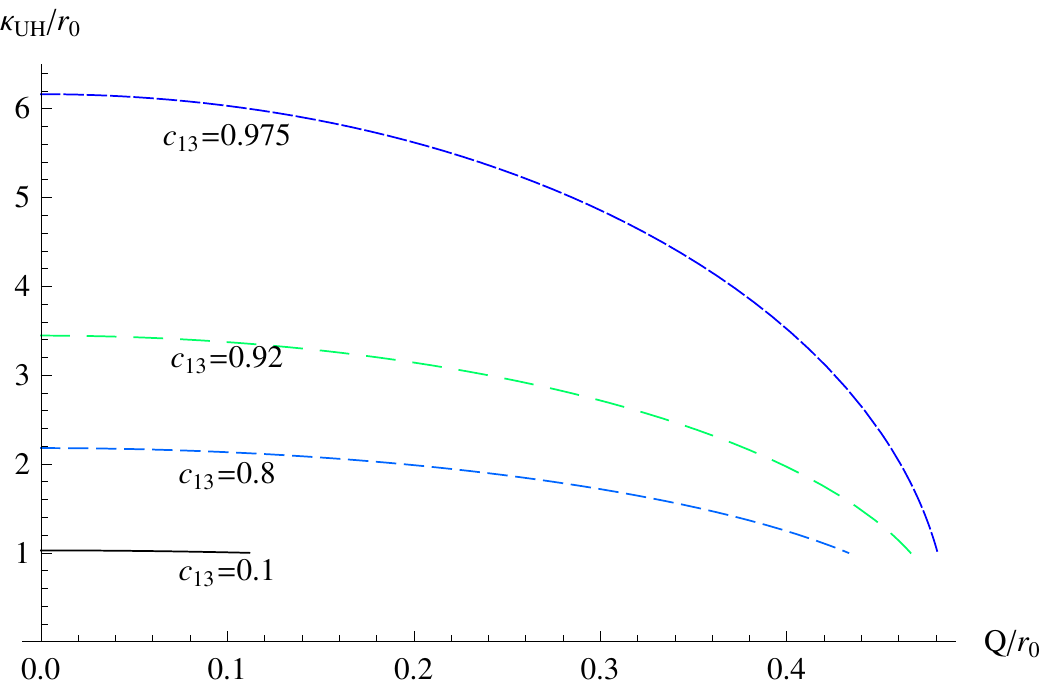}\;\;\;\;\;
\includegraphics[width=7.5cm]{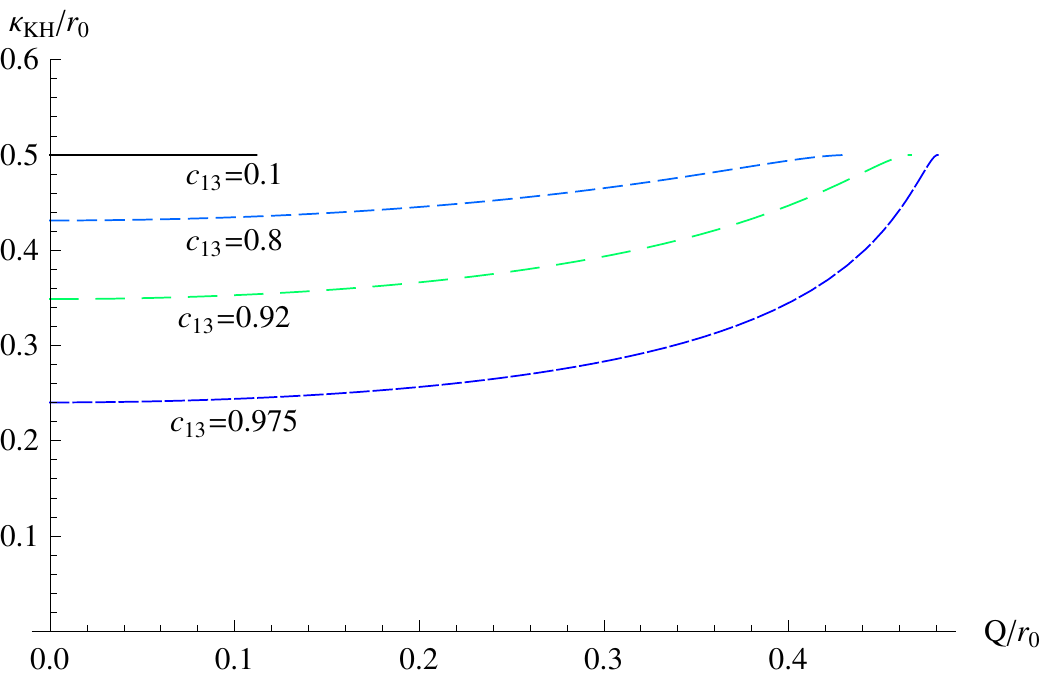}\;\;\;\;\;
\caption{The surface gravity at the universal horizon and Killing horizon of charged Einstein-aether black holes in the case of $c_{14}=0.1,~c_{123}=0$. The restrictions are $c_{13}\geq0.05,~Q/r_0\leq\sqrt{(2c_{13}-0.1)/8}$.}
 \label{fig3}
  \end{figure}

The position of the universal horizon $r_{UH}$ and its surface  gravity (using Eq. \ref{kappa}) are
\begin{eqnarray}\label{kappauh2}
 r_{UH}=\frac{r_0}{2},\quad  \kappa_{UH}=\frac{1}{2\sqrt{(1-c_{13})}r_{UH}}
\sqrt{1-\frac{c_{14}}{2}-\frac{Q^2}{r_{UH}^2}}.
\end{eqnarray}
And the $\kappa_{UH}$ is showed in Figure \ref{fig3}.
Also, when $Q=0$ it reduces to the one obtained  in  \cite{cropp,berglund2}.
One can see that $r_{UH}$  does not depend on the charge $Q$, but $\kappa_{UH}$ depends on it.
In other words, to the same universal horizon $r_{UH}= r_0/2$, there are different thermal temperatures of the horizon with different charge $Q$, if we assume that
$T$ is still somehow proportional to $\kappa_{UH}$.

The Smarr formula new reads,
\begin{eqnarray}\label{quh2}
 &&G_{\ae} M=\frac{q_{UH}A_{UH}}{4\pi }+V_{UH}Q,\quad q_{UH}=\frac{1}{ r_{UH}}\left(1-\frac{c_{14}}{2}-\frac{Q^2}{r_{UH}^2}\right),\nonumber\\
   &&G_{\ae} \delta M=\Big(1-\frac{c_{14}}{2}\Big)\frac{1}{r_{UH}}\frac{\delta A_{UH}}{8\pi},
\end{eqnarray}
in which  the  term proportional to $\delta Q$ is absent. 
To see this more clearly, let us
consider the Smarr method, from which we find that
\begin{eqnarray}
\label{delta1a2}
  G_{\ae} M=\left(1-\frac{c_{14}}{2}\right)\sqrt{\frac{A}{4\pi}}.
\end{eqnarray}
Then,  writing the variation of $M$ as $G_{\ae} \delta M=T\delta A+V\delta Q$, we  obtain,
\begin{eqnarray}
\label{delta22}
  T=\frac{\partial \left(G_{\ae} M\right)}{\partial A}=\left(1-\frac{c_{14}}{2}\right)\frac{1}{8\pi r_{UH}},\quad V=\frac{\partial  \left(G_{\ae} M\right)}{\partial Q}=0,
\end{eqnarray}
which are the same as those given in the second line of Eq.(\ref{quh2}).
Once again,  $q_{UH}$ and $T$ aren't proportional to the $\kappa_{UH}$ given by Eq.(\ref{kappauh2}).

On the other hand, the Killing horizon and  its surface  gravity (using Eq. \ref{kappa1}) are
\begin{eqnarray}\label{rkh2}
 r_{KH}=r_0+r_u,\quad  \kappa_{KH}=\frac{2r_u+r_0}{2r_{KH}^2},
\end{eqnarray}
which are showed in Figure \ref{fig3}. In order for them to be real we must assume that $Q\leq \sqrt{1-\frac{c_{14}}{2}}\frac{r_0}{2}$.
Comparing Eq.(\ref{alruh}) with (\ref{rkh2}), one can see that the latter condition on $Q$ is contained in the former, i.e., in
the charged aether black hole, the  electric charge is subjected to more stringent restrictions. Again when $c_{13}=c_{14}=0$,  $\kappa_{KH}$ reduces to $(r_+-r_-)/2r_+^2$. %
The Smarr formula  at the Killing horizon is
\begin{eqnarray}\label{qkh2}
 &&G_{\ae} M=\frac{q_{KH}A_{KH}}{4\pi }+V_{KH}Q,\quad q_{KH}=\left[\left(1-\frac{c_{14}}{2}\right)\frac{r_0}{2r_{KH}^2}
 -\frac{Q^2}{r_{KH}^3}\right],
\end{eqnarray}
which depends  on the coupling constants $c_{13}$ and $c_{14}$.  $q_{KH}$ approaches to $(r_+-r_-)/2r_+^2$,
 if $c_{13}=c_{14}=0$. We find that taking variation
with respect to each term  cannot obtain  the first law, so instead  we use Smarr's method \cite{smarr}, and find that
\begin{eqnarray}
  &&G_{\ae}\delta M=\frac{\partial M}{\partial A}\delta A_{KH}+\frac{\partial M}{\partial Q}\delta Q,
   \quad \frac{\partial M}{\partial Q}=\frac{c_ac_bQ}{\sqrt{c_a(c_ac_b-1)Q^2+c_br_{KH}^2}},  \quad c_a\equiv \frac{1}{1-c_{13}},
  \nonumber\\
  && c_b\equiv1-\frac{c_{14}}{2}, \quad
 T=\frac{\partial M}{\partial A}=
  \frac{c_b}{c_ac_b-1}\left(\frac{c_ac_b}{\sqrt{c_a(c_ac_b-1)Q^2+c_ac_br_{KH}
  ^2}}-\frac{1}{r_{KH}}\right).
\end{eqnarray}
 {Note that, similar to the previous case,  now $q_{KH}$ and $T$ aren't proportional to the $\kappa_{KH}$ given by Eq.(\ref{rkh2}), either.}

Finally, from Figure \ref{fig3} we note that
the dependence of $\kappa_{UH}$ on $Q$ is similar to the former
 case. In particular,
 its presence always makes the temperature lower, while  the presence of
$c_{13}$ increases it. The surface gravity $\kappa_{UH}$  is always larger than
$\kappa_{KH}$. At the Killing horizon, the effects of the charge and $c_{13}$ on $\kappa_{KH}$ are just
opposite.

\section{Conclusions}

In this paper, we have studied  the Einstein-Maxwell-aether theory, and  found two new classes  of charged black hole solutions
for the special choices of the coupling constants: (1) $c_{14}=0,~c_{123}\neq0$, and (2) $c_{14}\neq0,~c_{123}=0$. In the first case, the universal horizon depends on its electric charge $Q$, while it  doesn't in the second case.
In both of the cases, the  universal horizons are independent of the coupling constants $c_i$, while the
Killing horizons depend on them. When $c_{13}\;  (\equiv c_1 + c_3)$ is very small and approaches to zero, the solutions in the case $c_{14}=0,~c_{123}\neq0$ reduce to the usual Reissner-Nordstrom black hole solution. The corresponding properties at the universal horizons are the same as those presented  in \cite{lin} via Khronometric theory.

To study the solutions further, we have considered their surface gravity and constructed the Smarr formula
at each of the  horizons, universal and Killing. We have shown that there is no problem for such constructions, but when  trying  to construct the
corresponding  first law of black hole mechanics, they  are all different
 { from the usual one.} In particular, we have shown that the temperature obtained from the  Smarr mass-area relation is not proportional to its corresponding surface gravity,
 when both of the charge and aether are present, in contrast to the case without aether ($c_i = 0$) \cite{lin}, or to the case without charge \cite{berglund}.
In particular,
in \cite{berglund}  it was found that  in the neutral case  $q_{UH}$ is always proportional to the  surface gravity $\kappa_{UH}$ at the  universal horizons,
even when the aether is present. In the present paper, from Eqs.(\ref{qkh1kappakh1}), (\ref{rkh2}) and (\ref{qkh2})
we can see that, when $Q=0$, $q_{KH}$ is also proportional to $\kappa_{KH}$. Then, one can also be able to construct  a slightly modified
first law of black hole mechanics at  the  Killing  horizons.

However, when the charge $Q$ is different from zero, comparing
$(q_{UH}, q_{KH})$ with $(\kappa_{UH}, \kappa_{KH})$, one can see that
 these proportional relations don't hold any longer.
It was also found that in the presence of the cosmological constant\cite{bhattacharyya}, $q_{UH}$ is also not proportional to $\kappa_{UH}$.
Therefore, it is not clear how to build the first law for
these charged aether black holes, before we have a better understanding of the entropy of the universal and/or Killing
horizons.

The solutions presented in this paper can be generalized to the case coupled with the cosmological constant $\Lambda$, which  are given by
\begin{eqnarray}
\label{CSolutions}
  e(r)=\left\{
\begin{array}{c}
  1-\frac{r_0l_s+2r_{\ae}^2}{l_sr}+\frac{Q^2}{r^2}-c_{13}\frac{r_{\ae}^4}{r^4}
 +(\frac{2c_2+c_{123}}{2l_s^2}-\frac{1}{3}\Lambda) r^2,\quad (c_{14}=0,~c_{123}\neq0),
 \\ 1-\frac{r_0}{r}+\frac{c_{14}-2c_{13}}{2(1-c_{13})}\frac{r_0^2}{4r^2}
 +\frac{1}{1-c_{13}}(\frac{Q^2}{r^2}-\frac{1}{3}\Lambda r^2)+\frac{r_s}{r},\quad (c_{14}\neq0,~c_{123}=0),
\end{array}
\right.
\end{eqnarray}
where $r_0, \; r_{\ae},\; l_s$ and $r_s$ are integration constants.

In addition,  from these solutions, one can also   construct topological charged
Einstein-aether (anti) de Sitter  black holes, which are
\begin{equation}
ds^2 = -\big[e(r) - 1\big] dv^2 + 2dv dr + r^2\left(d\hat\theta^2 + d\hat\phi^2\right),
\end{equation}
where $e(r)$ is given by Eq.(\ref{CSolutions}).  The studies of the properties of the above solutions are out of the scope of this paper, and we
hope to report them  in another occasion soon.

\begin{acknowledgments}
We would like to thank Ted Jacobson for useful suggestions and comments. C.D.  was supported by NNSFC No. 11247013, Hunan Provincial NSFC No. 2015JJ2085, State Scholarship Fund of China No. 201308430340, and the Aid program for Science and Technology Innovative Research Team in Higher Educational Institutions of Hunan Province. A.W. was supported
in part by Ci\^{e}ncia Sem Fronteiras, No.
A045/2013 CAPES, Brazil and NNSFC No.
11375153, China. This work was done during
C.D.'s  visiting at Baylor University and he would like
to thank Baylor for hospitality.
\end{acknowledgments}

\end{document}